\documentclass[twocolumn,letter]{jpsj3}
\usepackage{txfonts}
\RequirePackage{graphicx,color}

\title{
Magnetoacoustic Resonance to Probe Quadrupole--Strain Coupling \\
in a Diamond Nitrogen-Vacancy Center as a Spin-Triplet System
}

\author{Mikito Koga$^1$ and Masashige Matsumoto$^2$}

\inst{
$^1$Department of Physics, Faculty of Education, Shizuoka University, Shizuoka
422--8529, Japan \\
$^2$Department of Physics, Faculty of Science, Shizuoka University, Shizuoka
422--8529, Japan \\
}

\abst{
A theory of magnetoacoustic resonance is proposed to measure
quadrupole--strain couplings in a spin-triplet state with the $C_{3v}$ point group symmetry,
considering the spin--strain interaction in a diamond nitrogen-vacancy (NV) center.
Based on the Floquet theory, we demonstrate how the
single- and two-phonon transition probabilities depend on the change in the longitudinal and
transverse quadrupole couplings, which can be controlled by rotating an applied magnetic field, 
around the threefold axis.
The obtained quadrupole dynamics results are useful for realizing mechanical or ac strain-control
of the NV spin as an alternative to the conventional magnetic control by spin resonance.
}

\begin{document}

\maketitle

\newcommand{\ds}{\displaystyle}

\renewcommand{\H}{\mathcal{H}}
\newcommand{\br}{{\mbox{\boldmath$r$}}}
\newcommand{\bR}{{\mbox{\boldmath$R$}}}
\newcommand{\bS}{{\mbox{\boldmath$S$}}}
\newcommand{\bk}{{\mbox{\boldmath$k$}}}
\newcommand{\bH}{{\mbox{\boldmath$H$}}}
\newcommand{\bh}{{\mbox{\boldmath$h$}}}
\newcommand{\bJ}{{\mbox{\boldmath$J$}}}
\newcommand{\bI}{{\mbox{\boldmath$I$}}}
\newcommand{\bPsi}{{\mbox{\boldmath$\Psi$}}}
\newcommand{\bpsi}{{\mbox{\boldmath$\psi$}}}
\newcommand{\bPhi}{{\mbox{\boldmath$\Phi$}}}
\newcommand{\bd}{{\mbox{\boldmath$d$}}}
\newcommand{\bG}{{\mbox{\boldmath$G$}}}
\newcommand{\bu}{{\mbox{\boldmath$u$}}}
\newcommand{\be}{{\mbox{\boldmath$e$}}}
\newcommand{\om}{{\omega_n}}
\newcommand{\omm}{{\omega_{n'}}}
\newcommand{\omd}{{\omega^2_n}}
\newcommand{\omt}{{\tilde{\omega}_{n}}}
\newcommand{\ommt}{{\tilde{\omega}_{n'}}}
\newcommand{\btau}{{\hat{\tau}}}
\newcommand{\brho}{{\mbox{\boldmath$\rho$}}}
\newcommand{\bsigma}{{\mbox{\boldmath$\sigma$}}}
\newcommand{\bSigma}{{\mbox{\boldmath$\Sigma$}}}
\newcommand{\bt}{{\hat{t}}}
\newcommand{\bq}{{\hat{q}}}
\newcommand{\bLambda}{{\hat{\Lambda}}}
\newcommand{\bDelta}{{\hat{\Delta}}}
\newcommand{\bU}{{\hat{U}}}
\newcommand{\bskp}{{\mbox{\scriptsize\boldmath $k$}}}
\newcommand{\skp}{{\mbox{\scriptsize $k$}}}
\newcommand{\bsrp}{{\mbox{\scriptsize\boldmath $r$}}}
\newcommand{\bsRp}{{\mbox{\scriptsize\boldmath $R$}}}
\newcommand{\bsk}{\bskp}
\newcommand{\sk}{\skp}
\newcommand{\bsr}{\bsrp}
\newcommand{\bsR}{\bsRp}
\newcommand{\ri}{{\rm i}}
\newcommand{\re}{{\rm e}}
\newcommand{\rd}{{\rm d}}
\newcommand{\rM}{{\rm M}}
\newcommand{\rs}{{\rm s}}
\newcommand{\rt}{{\rm t}}
\newcommand{\Tc}{{$T_{\rm c}$}}
The negatively charged nitrogen-vacancy (NV) center in diamond is a unique defect in
which the spin degrees of freedom are described by spin-1 ($S = 1$) in a $C_{3v}$
crystalline-electric-field environment.
\cite{Lenef96,Goss96,Gali08,Doherty11}
A long coherence time of over a millisecond is a significant advantage for the robustness of the spin state
at room temperature.
\cite{Balasubramanian09,Mizuochi09,Herbschleb19}
Thus, the NV center is a good candidate for a promising platform for spin-controlled
devices for quantum information processing and sensing applications.
\cite{Suter17}
Since the $S=1$ spin operator $\bS = (S_x, S_y, S_z)$ contains quadrupole degrees of freedom,
the electronic spin is coupled to local strains owing to the crystal lattice deformations.
There are five components of quadrupole operators: 
$O_u = ( 2 S_z^2 - S_x^2 - S_y^2 ) / \sqrt{3}$,
$O_v = S_x^2 - S_y^2$, $O_{xy} = S_x S_y + S_y S_x$,
$O_{zx} = S_z S_x + S_x S_z$, and $O_{yz} = S_y S_z + S_z S_y$.
\par

Recently, a theoretical proposal for mechanically and electrically driven electron spin resonance
has shed light on the important role of spin--strain interaction in the spin-triplet ground state of
the NV defect,
\cite{Udvarhelyi18}
which is split into lower singlet ($S_z = 0$) and higher doublet ($S_z = \pm 1$) energy levels
by a uniaxial crystal field along the threefold axis.
This work has pointed out the relevance of $O_{zx}$ and $O_{yz}$ to
electrical or mechanical control of the NV spin, although not much attention has been paid to these
quadrupoles to date.
\cite{Kiel72,Mims76,Oort90,Doherty12,Doherty13}
Since only $O_u$, $O_v$, and $O_{xy}$ have been considered for such spin control,
\cite{MacQuarrie13,Klimov14,Barfuss15}
the confirmation of $O_{zx}$ and $O_{yz}$ is highly desired for pursuit of various methods of electrical
or mechanical spin control as an alternative to conventional magnetic control.
\cite{Lee17,Barfuss19}
Note that $O_u$, $O_v$, and $O_{xy}$ cause the transition in the doublet, whereas
$O_{zx}$ and $O_{yz}$ are involved in the transition between the singlet and doublet levels.
Very recently, an evaluation of spin--strain coupling with $O_{zx}$ has been performed by
measurements of an acoustically driven single-quantum spin transition.
\cite{Chen20}
As reported in our recent studies,
\cite{Koga20,Matsumoto20}
it is also important that the transition via quadrupole couplings can be changed by rotating an
applied magnetic field.
This is useful for probing such a spin--strain coupling with $O_{zx}$ that is difficult to measure.
\par

In this study, we present a new idea of magnetoacoustic resonance for ultrasonic measurements
of spin--strain coupling parameters in the $S=1$ spin state, considering the NV spin
as a typical example.
This was first motivated by the discovery of an extremely strong strain coupling inherent in boron-doped silicon vacancies by elastic softening measurements.
\cite{Goto06,Mitsumoto14}
In the NV center, phonon-assisted orbital transitions driven by an acoustic wave were detected by photoluminescence excitation spectroscopy
\cite{Chen18}
as well as an acoustically driven transition in the spin-triplet state.
\cite{Chen20}
Thus, the vacancy states with quadrupoles commonly possess high sensitivity to local strains or
lattice vibrations.
\par

We study a simplified spin--strain interaction in the electronic $S = 1$ spin state,
considering that the lattice deformations are limited in the plane including the $[001]$ and $[110]$
crystal axes.
Using the above quadrupole operators in the $C_{3v}$ frame (NV axis frame), the spin--strain
interaction Hamiltonian can be written as
\cite{Udvarhelyi18,Suppl1}
\begin{align}
H_\varepsilon = \sum_k A_{k, \varepsilon} O_k~~(k = u, v, zx),
\label{eqn:Hep}
\end{align}
where $A_{k, \varepsilon}$ is a strain-dependent coupling coefficient with each quadrupole, and
both $A_{xy, \varepsilon}$ and $A_{yz, \varepsilon}$ vanish owing to the limited lattice deformations.
The $z$-axis is chosen in the direction of a threefold axis vector $\be_z = (1,1,1)/\sqrt{3}$, and the
two other orthogonal basis vectors are defined as $\be_y = (1,-1,0)/\sqrt{2}$ and
$\be_x = (-1,-1,2)/\sqrt{6}$.
The coupling coefficients are given by
$A_{u, \varepsilon} = g_a \varepsilon_1$,
$A_{v, \varepsilon} = ( g_b \varepsilon_{U_1} + g_c \varepsilon_{U_2}  ) / \sqrt{3}$, and
$A_{zx, \varepsilon} = ( 2 g_d \varepsilon_{U_1} - g_e \varepsilon_{U_2} ) / \sqrt{6}$,
where
$\varepsilon_1 = ( \varepsilon_{YZ} + \varepsilon_{ZX} + \varepsilon_{XY} ) / \sqrt{3}$,
$\varepsilon_{U_1} = (2\varepsilon_{ZZ} - \varepsilon_{XX} - \varepsilon_{YY}) / \sqrt{3}$, and
$\varepsilon_{U_2} = (2\varepsilon_{XY} - \varepsilon_{YZ} - \varepsilon_{ZX}) / \sqrt{3}$.
The strain tensors are denoted by
$\varepsilon_{ij} = [ ( \partial u_i / \partial x_j ) + ( \partial u_j / \partial x_i ) ] / 2$
with the displacement vector $\bu = (u_1, u_2, u_3) = (u_X, u_Y, u_Z)$, and
$(x_1, x_2, x_3) = (X,Y,Z)$ is the cubic crystal coordinate.
There are five independent coupling parameters $g_i$ ($i = a, b, c, d, e$), and bulk strain
$\varepsilon_{XX} + \varepsilon_{YY} + \varepsilon_{ZZ}$ has been disregarded.
\par

In the $C_{3v}$ $(xyz)$ frame, the electronic $S = 1$ states are described by the
following local Hamiltonian:
\begin{align}
H_{\rm l} = - h ( S_x \cos \phi + S_y \sin \phi ) + \sqrt{3} D O_u,
\end{align}
where $h = \gamma_{\rm e} H$ for the magnetic field $\bH = ( H \cos \phi, H \sin \phi, 0)$ 
($\gamma_{\rm e} = 2.8$ MHz/G is the electron gyromagnetic ratio).
In the last term, $3D$ ($>0$) equals the energy of the doublet excited state measured from the
singlet ground state for $H = 0$ and this splitting is $2.87$ GHz for the NV center.
The doublet state is split by the magnetic field, and we neglect the higher-lying state
assuming that $h$ is sufficiently large compared to $A_{k, \varepsilon}$ in Eq.~(\ref{eqn:Hep}).
Note that the energies of the three spin states do not depend on the field direction $\phi$
perpendicular to the threefold axis.
After diagonalizing $H_{\rm l}$, the eigenvalue and eigenfunction of the ground state are obtained
as
\begin{align}
& \frac{E_1}{D} = \frac{1}{2} ( -1 - \alpha )~~
\left( \alpha = \sqrt{ 9 + 4 \bar{h}^2 },~~\bar{h} = \frac{h}{D} \right), \\
& | \psi_1 \rangle = \frac{ \sin \chi }{ \sqrt{2} } e^{ - i \phi } | + 1 \rangle + \cos \chi | 0 \rangle
 + \frac{ \sin \chi }{ \sqrt{2} } e^{ i \phi } | - 1 \rangle,
\label{eqn:psi1}
\end{align}
respectively, based on the eigenstates $| m \rangle$ ($m = 0, \pm 1$) of $S_z$.
For the first excited state, we obtain
\begin{align}
\frac{E_2}{D} = 1,~~
| \psi_2 \rangle = \frac{1}{ \sqrt{2} } e^{ - i \phi } | + 1 \rangle - \frac{1}{ \sqrt{2} } e^{ i \phi } | - 1 \rangle.
\end{align}
The coefficients in Eq.~(\ref{eqn:psi1}) are given by
\begin{align}
\cos \chi = \sqrt{ \frac{1}{2} \left( 1 + \frac{3}{ 2 \bar{\varepsilon}_0 - 3} \right) },~~
\sin \chi = \sqrt{ \frac{1}{2} \left( 1 - \frac{3}{ 2 \bar{\varepsilon}_0 - 3} \right) },
\label{eqn:chi}
\end{align}
where $\bar{\varepsilon}_0 = ( E_2 - E_1 ) / D = ( 3 + \alpha ) / 2$.
Since $\bar{\varepsilon}_0 > 3$ must be satisfied, $\chi$ varies in $0 < \chi < \pi / 4$.
\par

Next, we derive an effective spin--strain interaction Hamiltonian in the subspace of the above
two states $| \psi_\mu \rangle$ ($\mu = 1,2$) coupled to time-dependent oscillating strain fields
$\varepsilon_\lambda$ ($\lambda = 1, U_1, U_2$), which are driven by an acoustic wave
propagating in the lattice.
The time dependency is represented by
$\varepsilon_\lambda = a_\lambda \cos \omega t$,
where $\omega$ is the acoustic-wave frequency and $a_\lambda$ is the vibration amplitude.
The relative phase shifts between the three components are not considered for simplicity.
By calculating $\langle \psi_\mu | H_\varepsilon | \psi_\nu \rangle$ ($\mu, \nu = 1,2$), we obtain the
following form of the effective Hamiltonian for the two-level system coupled to the periodically
time-dependent strains,
\begin{align}
H_{\rm eff} (t) = \frac{1}{2}
\left(
\begin{array}{cc}
- \varepsilon_0 - A_L \cos \omega t & A_T \cos \omega t \\
A_T^* \cos \omega t & \varepsilon_0 + A_L \cos \omega t
\end{array}
\right).
\label{eqn:Ht}
\end{align}
Here, $\varepsilon_0 = \bar{\varepsilon}_0 D$ is the level splitting of the two states.
The longitudinal
($A_L \cos \omega t = \langle \psi_2 | H_\varepsilon | \psi_2 \rangle
- \langle \psi_1 | H_\varepsilon | \psi_1 \rangle$) and transverse
($A_T \cos \omega t =  2 \langle \psi_1 | H_\varepsilon | \psi_2 \rangle$)
couplings depend on the magnetic field direction $\phi$ and the $\bar{\varepsilon}_0$-dependent
trigonometric functions in Eq.~(\ref{eqn:chi}),
\begin{align}
& A_L = \frac{\sqrt{3}}{2} ( 1 + \cos 2 \chi ) A_u - \frac{1}{2} ( 3 - \cos 2 \chi ) \cos 2 \phi \cdot A_v,
\label{eqn:ALtwo} \\
& A_T = 2 ( - i \sin \chi \sin 2 \phi \cdot A_v + \cos \chi \cos \phi \cdot A_{zx} ).
\label{eqn:ATtwo}
\end{align}
For the quadrupole--strain couplings, $A_u = g_a a_1$,
$A_v = ( g_b a_{U_1} + g_c a_{U_2}  ) / \sqrt{3}$, and
$A_{zx} = ( 2 g_d a_{U_1} - g_e a_{U_2} ) / \sqrt{6}$.
\par

Similar forms of the time-dependent Hamiltonian in Eq.~(\ref{eqn:Ht}) has been frequently studied
by the Floquet theory.
\cite{Koga20,Matsumoto20,Shirley65,Chu04,Son09,Hausinger10}
Following Shirley,
\cite{Shirley65}
the problem of solving the time-dependent
Schr\"{o}dinger equation is transformed to a time-independent eigenvalue problem
using an infinite-dimensional matrix form of the Floquet Hamiltonian.
The matrix element $\langle \alpha n | H_F | \beta m \rangle
= H_{\alpha \beta}^{ [ n - m ] } + n \omega \delta_{nm} \delta_{\alpha \beta}$
is constructed using the Floquet states
$| \alpha n \rangle = | \alpha \rangle \otimes | n \rangle$.
Here, $\alpha$ ($= \psi_1, \psi_2$) and $n$
($= 0, \pm 1, \pm2, \cdots$) denote one of the two levels and the time dependency
$e^{i n \omega t}$, respectively, and $\hbar = 1$ is used.
The block matrix $H^{[ n - m]}$ is only finite for $n - m = 0, \pm1$.
In the Floquet matrix represented by
\begin{align}
H_F =
\left(
\begin{array}{ccccccc}
\ddots & ~ & ~ & \vdots & ~ & ~ & ~ \\
~ & H_{-2}^{[0]} & H^{[-1]} & {\bf 0} & {\bf 0} & {\bf 0} & ~ \\
~ & H^{[1]} & H_{-1}^{[0]} & H^{[-1]} & {\bf 0} & {\bf 0} & ~ \\
\cdots & {\bf 0} & H^{[1]} & H_0^{[0]} & H^{[-1]} & {\bf 0} & \cdots \\
~ & {\bf 0} & {\bf 0} & H^{[1]} & H_1^{[0]} & H^{[-1]} & ~ \\
~ & {\bf 0} & {\bf 0} & {\bf 0} & H^{[1]} & H_2^{[0]} & ~ \\
~ & ~ & ~ & \vdots & ~ & ~ & \ddots
\end{array}
\right),
\label{eqn:HFblock}
\end{align}
the diagonal sectors are defined as
\begin{align}
H_n^{[0]} \equiv H^{[0]} + n \omega =
\left(
\begin{array}{cc}
- ( \varepsilon_0 / 2 ) + n \omega & 0 \\
0 & ( \varepsilon_0 / 2 ) + n \omega
\end{array}
\right),
\end{align}
and the off-diagonal sectors are given by
\begin{align}
H^{[ \pm 1 ]} = \frac{1}{4}
\left(
\begin{array}{cc}
- A_L & A_T \\
A_T^* & A_L
\end{array}
\right).
\end{align}
In Eq.~(\ref{eqn:HFblock}), $\bf 0$ represents the $2 \times 2$ form of the zero matrix.
The eigenvalue problem is described by, 
$H_F | q_\gamma \rangle = q_\gamma | q_\gamma \rangle$,
where the $\gamma$th eigenvalue $q_\gamma$ is termed the quasienergy and
$| q_\gamma \rangle$ is the corresponding eigenfunction.
For the time-averaged transition probability between the $\alpha$ and $\beta$ states, we use
the following formula:
$\bar{P}_{\alpha \rightarrow \beta} = \sum_m \sum_\gamma
| \langle \beta m | q_\gamma \rangle \langle q_\gamma | \alpha 0 \rangle |^2$.
\cite{Shirley65,Son09}
\par

In particular, we focus on the emergence of transition probability peaks expected for the nearly degenerate Floquet states, for instance, $| \alpha 0 \rangle$ and $| \beta, -n \rangle$, where
$ - \varepsilon_0 / 2 \simeq \varepsilon_0 / 2 - n \omega$ is satisfied.
In this case, the infinite-dimensional matrix form of $H_F$ is reduced to an effective $2 \times 2$
matrix using the Van Vleck perturbation theory.
\cite{Son09,Hausinger10}
By analogy with previous studies,
\cite{Koga20,Matsumoto20}
the effective Hamiltonian in the subspace of 
$| \alpha 0 \rangle$ and $| \beta, -n \rangle$ is given by
\begin{align}
\tilde{H}_F =
\left(
\begin{array}{cc}
- ( \varepsilon_0 / 2 ) + \delta_n & v_{-n} \\
v_{-n}^* & ( \varepsilon_0 / 2 ) - \delta_n - n \omega
\end{array}
\right).
\end{align}
Here, the off-diagonal matrix element
$v_{-n} = (- n \omega / 2) J_{-n} ( A_L / \omega ) A_T / A_L$
is calculated up to the first order of $A_T$ using the $k$th Bessel function of the
first kind $J_k$.
The leading term of the energy shift $\delta_n$ is given as
$\delta_n = - \sum_{k \ne - n} | v_k |^2 / ( \varepsilon_0 + k \omega )$.
The diagonalization of $\tilde{H}_F$ gives the eigenvalues
$q_\pm = - (n \omega)/2 \pm \tilde{q}_n$, where
$\tilde{q}_n = \sqrt{ ( n \omega - \varepsilon_0 + 2 \delta_n )^2 / 4 + | v_{-n} |^2 }$, and leads to the
time-averaged transition probability represented by
$\bar{P}_{\psi_1 \rightarrow \psi_2}^{(n)} = (1 / 2) | v_{-n} |^2 / \tilde{q}_n^2$.
This is valid for $n = 1$ and $n = 2$, which correspond to the single- and two-phonon transition
processes, respectively, because the higher-order terms with $A_T$ must be considered for
$n \ge 3$.
In the weak coupling limit ( $| A_L | / \omega, | A_T | / \omega \ll 1$ ),
$| v_{-n} | \simeq \{ | A_T | / [ 2^{n+1} ( n - 1 )! ] \} ( | A_L | / \omega )^{n-1}~(n \ge 1)$ leads to
simple analytic forms for the transition probability at fixed
$\varepsilon_0 = n \omega$ ($n = 1,2$) as
\cite{Matsumoto20}
\begin{align}
& \bar{P}_{\psi_1 \rightarrow \psi_2}^{(1)} (\varepsilon_0 = \omega)
= \frac{1}{2} \frac{1}{ 1 + [ | A_T | / (8 \omega) ]^2 }, \\
& \bar{P}_{\psi_1 \rightarrow \psi_2}^{(2)} (\varepsilon_0 = 2 \omega)
= \frac{1}{2} \frac{ 1 }{ 1 + ( 4 / 9 ) ( | A_T | / | A_L | )^2 },
\label{eqn:P2}
\end{align}
for finite $| A_T |$, and $\bar{P}_{\psi_1 \rightarrow \psi_2}^{(n)}$ vanishes for $ A_T = 0$.
\par

In Eq.~(\ref{eqn:P2}), the two-phonon transition probability strongly depends on
$| A_T |^2 / | A_L |^2$, and it vanishes, especially when $ A_L $ approaches zero.
It must be noted that the longitudinal ($A_L$) coupling is required for the two-phonon transition
process as well as the transverse ($A_T$) coupling, as pointed out in $S = 1/2$ spin systems.
\cite{Gromov00}
This is completely unlike the single-phonon transition process dominated by $A_T$.
Since the quadrupole--strain couplings $A_L$ and $A_T$ depend on the rotation angle $\phi$
of the magnetic field, $P^{(2)}$ in Eq.~(\ref{eqn:P2}) changes with $\phi$.
From Eqs.~(\ref{eqn:ALtwo}) and (\ref{eqn:ATtwo}), the $\phi$ dependence is given by
\begin{align}
\frac{ | A_T |^2 }{ | A_L |^2 } = \frac{16}{9}
\frac{ A_{zx}^2 \cos^2 \chi \cos^2 \phi + A_v^2 \sin^2 \chi \sin^2 2 \phi }
{ \left[ \ds{ \frac{ A_u }{ \sqrt{3} } ( 1 + \cos 2 \chi ) - A_v \left( 1 - \frac{ \cos 2 \chi }{3} \right)
 \cos 2 \phi } \right]^2 }.
\label{eqn:ATAL}
\end{align}
Here, $\chi$ is given by substituting $\bar{\varepsilon}_0 = 2 \omega / D$ in Eq.~(\ref{eqn:chi}) and
is independent of $\phi$.
The ratios between the three couplings $A_u$, $A_v$, and $A_{zx}$ can be evaluated by
$\bar{P}^{(2)}$ as a function of $\phi$ in Eqs.~(\ref{eqn:P2}) and (\ref{eqn:ATAL}).
There exist characteristic field directions $\phi_0$ at which $\bar{P}^{(2)} \rightarrow 0$, namely,
$A_L \rightarrow 0$.
The ratio $A_u / A_v$ is obtained from
\begin{align}
\cos 2 \phi_0 = \frac{ A_u }{ \sqrt{3} A_v} \frac{ 1 + \cos 2 \chi }{ 1 - (\cos 2 \chi) / 3 },
\label{eqn:phi0}
\end{align}
and the absolute value of the right-hand side must be less than unity.
This evaluation is also valid for a stronger coupling case
($| A_T | / \omega, |A_L| / \omega \simeq 1$), as discussed later, and
$\bar{P}^{(2)}$ shows a minimum at $\phi \simeq \phi_0$.
For $| A_u / A_v | > \sqrt{3}$, no minimum is found in $0 < \phi < \pi$. 
In addition, for the weak coupling, the ratio $A_{zx} / A_v$ is related to the value of
$\bar{P}^{(2)}$ at $\phi = 0$ as
\begin{align}
\bar{P}^{(2)} ( \phi = 0 ) = \frac{1}{2}
\left\{ 1 + \left[ \frac{8}{9} C( \chi, \phi_0 ) \frac{ A_{zx} }{ A_{v} } \right]^2 \right\}^{-1},
\label{eqn:p2phi0}
\end{align}
where $C( \chi, \phi_0 ) = \cos \chi / \{ [ 1 - ( \cos 2 \chi ) / 3 ] ( 1 - \cos 2 \phi_0 ) \}$.
Thus, the ratios between the quadrupole couplings can be probed by
the field angle dependence of $\bar{P}^{(2)} (\phi)$ at $\varepsilon_0 = 2 \omega$.
In particular, for $A_v \gg A_u, A_{zx}$, $\bar{P}^{(2)} (\phi)$ shows fourfold symmetry in
rotating the magnetic field around the $z$ axis, namely, $[111]$, and it continuously approaches zero
at $\phi = \pi / 4$ in $0 < \phi < \pi / 2$.
As given by Eq.~(\ref{eqn:phi0}), $\phi$ for $P^{(2)} \rightarrow 0$ shifts to a lower value from
$\pi / 4$ with an increase in $A_u$.
In addition, the value of $\bar{P}^{(2)} (\phi = 0)$ decreases from $1/2$ as $A_{zx} / A_v$
increases.
Such features become more prominent in the weak coupling case.
\par

\begin{figure}
\begin{center}
\includegraphics[width=6.5cm,clip]{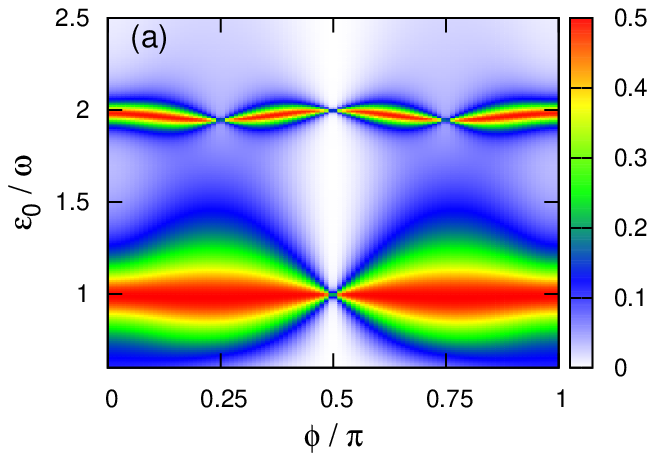}
\includegraphics[width=6.5cm,clip]{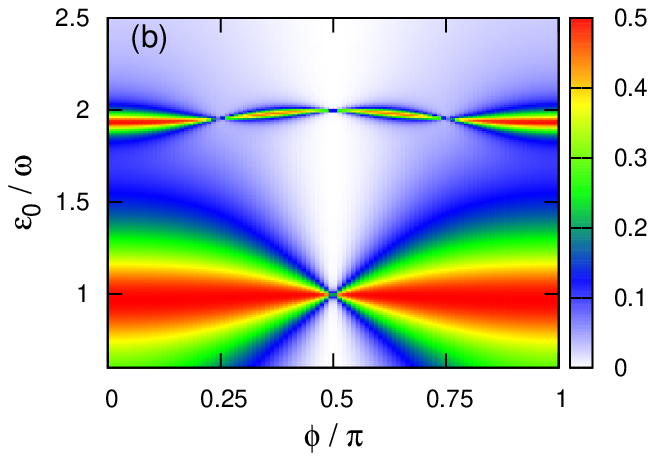}
\end{center}
\caption{
(Color online)
Contour map of the transition probability as a function of $\phi$ and $\varepsilon_0 / \omega$.
(a) $A_u = 0$, $A_v / \omega = 0.4$, and $A_{zx} / \omega = 0.2$.
(b) $A_u = 0$, $A_v / \omega = 0.2$, and $A_{zx} / \omega = 0.4$.
Here, $\varepsilon_0 / \omega > 0.6$ because $D / \omega$ is fixed at $0.2$.
}
\label{fig:1}
\end{figure}
First, let us consider the $A_u = 0$ case where the longitudinal coupling $A_L$ 
depends only on $A_v$.
Figures~\ref{fig:1} (a) and (b) show the contour maps of the transition probability $\bar{P}$
calculated numerically using the Floquet matrix, which are plotted as a function of
$\phi$ and $\varepsilon_0 / \omega$
[ $\varepsilon_0 = ( 3 D + \sqrt{ 9 D^2 + 4 h^2 } ) / 2$ ].
In both cases, $\bar{P}$ completely vanishes at $\phi / \pi = 1/2$ for all values
of $\varepsilon_0$ owing to $A_T = 0$ [see Eq.~(\ref{eqn:ATtwo})].
This indicates that the field direction $\phi / \pi = 1/2$ is very specific to the quadrupole--strain
coupling, which is parallel to $\be_y \parallel [1 \bar{1} 0]$ and perpendicular to the threefold axis.
The most prominent feature is the existence of a resonance peak at around
$\varepsilon_0 / \omega = 1$ associated with the single-phonon transition probability
$\bar{P}^{(1)}$.
In the weak coupling limit, the peak broadening $2 | v_{-1} |$ is proportional to $| A_T |$.
The $\phi$ dependence of $| A_T |$ explains the maximum broadening at $\phi / \pi \simeq 1/4$ for
$A_v > A_{zx}$ in Fig.~\ref{fig:1} (a),
which reflects $\sin 2 \phi$ for coupling with the $x^2 - y^2$ quadrupole
in Eq.~(\ref{eqn:ATtwo}).
\cite{Matsumoto20}
Conversely, the maximum broadening shifts toward $\phi = 0$ as $A_{zx} / A_v$ increases
in Fig.~\ref{fig:1} (b), owing to $\cos \phi$ for the coupling with the $zx$ quadrupole.
\par

To evaluate the spin--strain coupling parameters $g_i$, we focus on the
two-phonon transition process at around $\varepsilon_0 = 2 \omega$, although $\bar{P}^{(2)}$
shows a much narrower resonance peak in the weak coupling limit.
The two-phonon transition is dominated by longitudinal coupling with $A_L$.
$\bar{P} (\varepsilon_0 / \omega = 2)$ approaches the maximum $1/2$ for $| A_L | \gg | A_T | > 0$,
whereas it is strongly dependent on $| A_L |$, even when $| A_T | \simeq | A_L |$.
\par

\begin{figure}
\begin{center}
\includegraphics[width=6.5cm,clip]{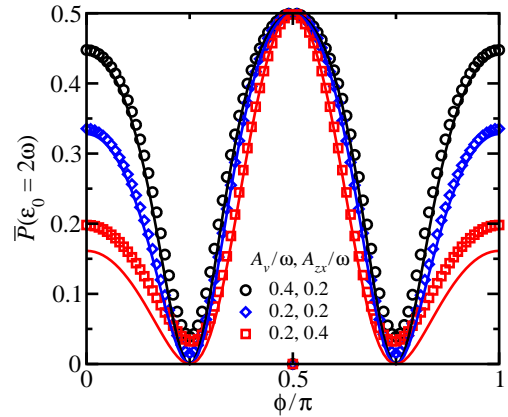}
\end{center}
\caption{
(Color online)
Transition probability at $\varepsilon_0 = 2 \omega$ as a function of $\phi$, where each data point is
plotted for $D / \omega =0.2$ and $A_u = 0$.
The black circles, blue diamonds, and red squares represent the data for
$(A_v, A_{zx}) = (0.4, 0.2)$, $(0.2, 0.2)$, and $(0.2, 0.4)$ in units of $\omega$, respectively.
The solid lines are drawn for $A_{zx} / A_v = 1/2$ (black), $1$ (blue), and $2$ (red) in the weak
coupling limit.
}
\label{fig:2}
\end{figure}
Figure~\ref{fig:2} shows $\bar{P} (\varepsilon_0 / \omega = 2)$ as a function of $\phi$
for various values of $A_v$ and $A_{zx}$, where $A_u$ is fixed at zero.
The ratio $A_{zx} / A_v$ is determined from the value at $\phi / \pi = 0$ or $1$ as a local maximum.
This value decreases with the increase in $A_{zx} / A_v$ as expected from Eq.~(\ref{eqn:p2phi0}).
For $A_{zx} / A_v > 1$ in Fig.~\ref{fig:2} (red squares), the relatively large deviations from the weak
coupling limit at around $\phi / \pi = 0$ and $1$ is owing to the larger contribution from higher order
terms with $A_{zx}$ compared to those with $A_v$.
For $A_u = 0$, the spin--strain coupling parameters $g_i$ can be evaluated by choosing a
quadrupole coupling with a single-strain component $\varepsilon_{U_1}$
($\varepsilon_1 = \varepsilon_{U_2} = 0$ for other strains), which provides $g_d / g_b = A_{zx} / (\sqrt{2} A_v)$.
\cite{Suppl2}
Although such an ideal measurement may be difficult for $A_u = 0$,
there are various methods of evaluating different coupling parameters (linear combinations of $g_i$),
which depend on the strain amplitudes $a_{U_1}$ and $a_{U_2}$
(See Sect.~S1.2, Supplemental Material).
\par

\begin{figure}
\begin{center}
\includegraphics[width=6.5cm,clip]{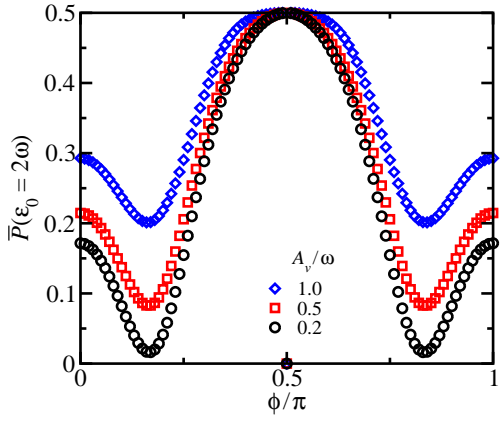}
\end{center}
\caption{
(Color online)
Transition probability at $\varepsilon_0 = 2 \omega$ as a function of $\phi$ for various values of
$A_v / \omega$, where $D / \omega = 0.2$ and $A_u / (\sqrt{3} A_v) = 0.4$, and $A_{zx} / A_v = 1$
are fixed.
The results for $A_v / \omega = 0.2$, $0.5$, and $1.0$ are plotted as circles, squares,
and diamonds, respectively.
}
\label{fig:3}
\end{figure}
For a finite $A_u / A_v$ ($>0$), $\bar{P} (\varepsilon_0 / \omega = 2)$ shows a minimum at
$\phi_0$ ($0 < \phi_0 / \pi < 1/4$ and $3/4 < \phi_0 / \pi < 1$) as plotted in Fig.~\ref{fig:3}, whereas
$\phi_0 / \pi = 1/4$ and $3/4$ for $A_u = 0$.
In the weak coupling limit, $\phi_0$ is specified by Eq.~(\ref{eqn:phi0}).
As long as $A_u / A_v$ is fixed at a constant, $\phi_0$ is not changed by increasing the three
couplings as we can see $\phi_0 / \pi = 1 / 6$ and $5/6$ in Fig.~\ref{fig:3}.
For instance, the ratio $g_a / g_c$ can be evaluated by $A_u / A_v$ when we choose
$\varepsilon_1 = \varepsilon_{U_2} / 2 = \varepsilon_{XY}$.
\cite{Suppl2}
For a single $\varepsilon_{U_2}$, $g_e / g_c$ ($= - \sqrt{2} A_{zx} / A_v$) is determined from the
value of $\bar{P} (\varepsilon_0 / \omega = 2)$ at $\phi = 0$ as mentioned above.
A similar analysis is also useful for evaluating $g_a / g_b$, $g_d / g_b$,
and various combinations of the spin--strain coupling parameters if the three strain
amplitudes $a_\lambda$ ($\lambda = 1, U_1, U_2$) can be adjusted.
On the experimental side of the NV center, unknown coupling parameters related to $A_{zx} / A_v$
have recently been measured using Rabi spectroscopy.
\cite{Chen20}
As a practical application of our theory, this challenging measurement can also be performed by
measuring the ultrasonic absorption rate, which is a widely used experimental method.
\par

In the present two-level system based on the $S=1$ spin with $C_{3v}$ symmetry, the
ground state $| \psi_1 \rangle$ is responsible for the magnetic moment
$M_\parallel = S_x \cos \phi + S_y \sin \phi$.
The time-averaged magnetic moment is obtained as
$\bar{M}_\parallel = ( 1 - \bar{P}_{\psi_1 \rightarrow \psi_2} ) \sin 2 \chi$, where $| \psi_1 \rangle$ is
chosen as the initial state.
Therefore, $\bar{M}_\parallel$ can be controlled by changing the magnetic field strength 
$h$ as well as the field direction $\phi$.
Note that $\bar{M}_\parallel$ shows no $\phi$ dependence when the lattice vibration is absent,
namely, no time-dependent strain is driven.
Using Eq.~(\ref{eqn:chi}), $\sin 2\chi = 2 \bar{h} / \sqrt{ 9 + 4 \bar{h}^2 }$ is obtained.
In particular, we focus on $\phi / \pi = 1/2$ in Fig.~\ref{fig:1}, where
$\bar{P}$ completely vanishes in the entire region of $\varepsilon_0 / \omega$ owing to $A_T = 0$.
When $\phi / \pi$ is tilted slightly from $1/2$, a finite transverse coupling generates an abrupt
increase in $\bar{P}$ at $\varepsilon_0 / \omega = 1, 2, 3, \cdots$, which leads to a sharp
resonance peak.
Consequently, $\bar{M}_\parallel$ shows an abrupt decrease in the field direction and shrinks
by half in the discrete fields
$h / \omega = \sqrt{ 1 - 3 D / \omega }$, $\sqrt{ 2 ( 2 - 3 D / \omega ) }$, $\cdots$.
This can also be realized by optical control using photon-assisted magnetoacoustic
resonance.
\cite{Koga20,Suppl3}
\par

In conclusion, we have demonstrated how the spin--strain coupling parameters for the $C_{3v}$
point group are revealed by the magnetoacoustic resonance, especially in the two-phonon
transition processes, considering an application to the spin states of the NV center.
This phonon transition strongly depends on the change in the longitudinal and transverse
quadrupole--strain couplings between the two levels, which can be controlled by rotating a
magnetic field around the threefold axis of a defect.
The present results provide useful information for high-frequency ultrasonic measurements
of quadrupole degrees of freedom inherent in the NV spin state and promote the development
of mechanically or ac strain-controlled spin devices.

\acknowledgment
This work was supported by JSPS KAKENHI Grant Number 17K05516.

\end{document}


\maketitle

\renewcommand{\figurename}{Fig. S \hspace{-0.25cm}}
\renewcommand{\theequation}{S\arabic{equation}}
\renewcommand{\thesection}{S\arabic{section}}
\newcommand{\ds}{\displaystyle}

\renewcommand{\H}{\mathcal{H}}
\newcommand{\br}{{\mbox{\boldmath$r$}}}
\newcommand{\bR}{{\mbox{\boldmath$R$}}}
\newcommand{\bS}{{\mbox{\boldmath$S$}}}
\newcommand{\bk}{{\mbox{\boldmath$k$}}}
\newcommand{\bH}{{\mbox{\boldmath$H$}}}
\newcommand{\bh}{{\mbox{\boldmath$h$}}}
\newcommand{\bJ}{{\mbox{\boldmath$J$}}}
\newcommand{\bI}{{\mbox{\boldmath$I$}}}
\newcommand{\bPsi}{{\mbox{\boldmath$\Psi$}}}
\newcommand{\bpsi}{{\mbox{\boldmath$\psi$}}}
\newcommand{\bPhi}{{\mbox{\boldmath$\Phi$}}}
\newcommand{\bd}{{\mbox{\boldmath$d$}}}
\newcommand{\bG}{{\mbox{\boldmath$G$}}}
\newcommand{\bu}{{\mbox{\boldmath$u$}}}
\newcommand{\be}{{\mbox{\boldmath$e$}}}
\newcommand{\om}{{\omega_n}}
\newcommand{\omm}{{\omega_{n'}}}
\newcommand{\omd}{{\omega^2_n}}
\newcommand{\omt}{{\tilde{\omega}_{n}}}
\newcommand{\ommt}{{\tilde{\omega}_{n'}}}
\newcommand{\btau}{{\hat{\tau}}}
\newcommand{\brho}{{\mbox{\boldmath$\rho$}}}
\newcommand{\bsigma}{{\mbox{\boldmath$\sigma$}}}
\newcommand{\bSigma}{{\mbox{\boldmath$\Sigma$}}}
\newcommand{\bt}{{\hat{t}}}
\newcommand{\bq}{{\hat{q}}}
\newcommand{\bLambda}{{\hat{\Lambda}}}
\newcommand{\bDelta}{{\hat{\Delta}}}
\newcommand{\bU}{{\hat{U}}}
\newcommand{\bskp}{{\mbox{\scriptsize\boldmath $k$}}}
\newcommand{\skp}{{\mbox{\scriptsize $k$}}}
\newcommand{\bsrp}{{\mbox{\scriptsize\boldmath $r$}}}
\newcommand{\bsRp}{{\mbox{\scriptsize\boldmath $R$}}}
\newcommand{\bsk}{\bskp}
\newcommand{\sk}{\skp}
\newcommand{\bsr}{\bsrp}
\newcommand{\bsR}{\bsRp}
\newcommand{\ri}{{\rm i}}
\newcommand{\re}{{\rm e}}
\newcommand{\rd}{{\rm d}}
\newcommand{\rM}{{\rm M}}
\newcommand{\rs}{{\rm s}}
\newcommand{\rt}{{\rm t}}
\newcommand{\Tc}{{$T_{\rm c}$}}
\renewcommand{\Pr}{{PrOs$_4$Sb$_{12}$}}
\newcommand{\La}{{LaOs$_4$Sb$_{12}$}}
\newcommand{\LaPr}{{(La$_{1-x}$Pr${_x}$)Os$_4$Sb$_{12}$}}
\newcommand{\PrLa}{{(Pr$_{1-x}$La${_x}$)Os$_4$Sb$_{12}$}}
\newcommand{\OsRu}{{Pr(Os$_{1-x}$Ru$_x$)$_4$Sb$_{12}$}}
\newcommand{\PrRu}{{PrRu$_4$Sb$_{12}$}}


\section{Spin--Strain Interaction for $S=1$ with $C_{3v}$ Symmetry}
\subsection{Spin--strain interaction Hamiltonian}
We start from the spin--strain interaction Hamiltonian for the spin-1 state with $C_{3v}$
symmetry and next derive a simplified Hamiltonian introducing some constraints to strain tensors.
Using the operators for the five quadrupole components expressed by the spin operator
$\bS = (S_x, S_y, S_z)$ for $S = 1$,
\begin{align}
& O_u = \frac{1}{\sqrt{3}} ( 2 S_z^2 - S_x^2 - S_y^2 ) = \frac{1}{\sqrt{3}} [ 3 S_z^2 - S (S + 1) ],
\nonumber \\
& O_v = S_x^2 - S_y^2,~~O_{xy} = S_x S_y + S_y S_x,
\nonumber \\
& O_{zx} = S_z S_x + S_x S_z,~~O_{yz} = S_y S_z + S_z S_y,
\label{eqn:Ok}
\end{align}
the spin--strain interaction Hamiltonian is written as
\begin{align}
H_\varepsilon = \sum_k A_{k, \varepsilon} O_k~~(k = u, v, xy, zx, yz),
\label{eqn:Hep}
\end{align}
where $A_{k, \varepsilon}$ is a strain-dependent coupling with each quadrupole $O_k$.
For $C_{3v}$ symmetry, the complete forms of $A_{k, \varepsilon}$ are given as
\cite{Udvarhelyi18}
\begin{align}
& A_{u, \varepsilon}
= \frac{1}{\sqrt{3}} [ h_{41} ( \varepsilon_{xx} + \varepsilon_{yy} ) + h_{43} \varepsilon_{zz} ],
\nonumber \\
& A_{v, \varepsilon} = - \frac{1}{2} \left[ h_{16} \varepsilon_{zx}
- \frac{1}{2} h_{15} ( \varepsilon_{xx} - \varepsilon_{yy} ) \right],
\nonumber \\
& A_{xy, \varepsilon} = \frac{1}{2} ( h_{16} \varepsilon_{yz} + h_{15} \varepsilon_{xy} ),
\nonumber \\
& A_{zx, \varepsilon} = \frac{1}{2} \left[ h_{26} \varepsilon_{zx}
 - \frac{1}{2} h_{25} ( \varepsilon_{xx} - \varepsilon_{yy} ) \right],
 \nonumber \\
& A_{yz, \varepsilon} = \frac{1}{2} ( h_{26} \varepsilon_{yz} + h_{25} \varepsilon_{xy} ).
\label{eqn:Ak}
\end{align}
Here, the three basis vectors in the $C_{3v}$ frame are chosen as $\be_x = (-1,-1,2)/\sqrt{6}$,
$\be_y = (1,-1,0)/\sqrt{2}$, and $\be_z = (1,1,1)/\sqrt{3}$.
Note that this choice of the $x$- and $y$-coordinates is different from the conventional manner.
For the latter, the $x$- and $y$-axes are parallel to the binary ($[1\bar{1}0]$) and
bisectrix ($[11\bar{2}]$) axes, respectively.
\begin{align}
\varepsilon_{ij} = \frac{1}{2} \left( \frac{\partial u_i}{\partial x_j}
 + \frac{\partial u_j}{\partial x_i} \right)
\label{eqn:epsilon}
\end{align}
denotes the strain tensor with the displacement vector $\bu = (u_x, u_y, u_z)$ and
$(x_1, x_2, x_3) = (x,y,z)$.
The $A_{u, \varepsilon} O_u$ term is equivalent to $\sqrt{3} A_{u, \varepsilon} S_z^2$ except for a
common energy shift of the spin states.
The spin--strain couplings are characterized by the six independent real parameters
$h_{41}$, $h_{43}$, $h_{15}$, $h_{16}$, $h_{25}$, and $h_{26}$.
\cite{Udvarhelyi18}
A derivation of $H_\varepsilon$ will be shown in Sect.~S3.
\par

Next, we transform the strain components in the $C_{3v}$ ($xyz$) coordinate to those of
the cubic crystal ($XYZ$) coordinate.
For the latter, $X \parallel [100]$, $Y \parallel [010]$, and $Z \parallel [001]$.
The relationship between the two coordinates is as follows:
\begin{align}
& \varepsilon_{xx} + \varepsilon_{yy} + \varepsilon_{zz}
= \varepsilon_{XX} + \varepsilon_{YY} + \varepsilon_{ZZ} \equiv \varepsilon_{\rm B},
\nonumber \\
& 2 \varepsilon_{zz} - \varepsilon_{xx} - \varepsilon_{yy}
= 2 (\varepsilon_{YZ} + \varepsilon_{ZX} + \varepsilon_{XY}),
\nonumber \\
& \varepsilon_{xx} - \varepsilon_{yy}
= \frac{1}{3} (2\varepsilon_{ZZ} - \varepsilon_{XX} - \varepsilon_{YY})
+ \frac{2}{3} (2\varepsilon_{XY} - \varepsilon_{YZ} - \varepsilon_{ZX}),
\nonumber \\
& \varepsilon_{zx}
= \frac{1}{3\sqrt{2}} (2\varepsilon_{ZZ} - \varepsilon_{XX} - \varepsilon_{YY})
- \frac{1}{3\sqrt{2}} (2\varepsilon_{XY} - \varepsilon_{YZ} - \varepsilon_{ZX}),
\nonumber \\
& \varepsilon_{xy}
= - \frac{1}{2\sqrt{3}} (\varepsilon_{XX} - \varepsilon_{YY})
- \frac{1}{\sqrt{3}} (\varepsilon_{YZ} - \varepsilon_{ZX}),
\nonumber \\
& \varepsilon_{yz}
= \frac{1}{\sqrt{6}} (\varepsilon_{XX} - \varepsilon_{YY})
- \frac{1}{\sqrt{6}} (\varepsilon_{YZ} - \varepsilon_{ZX}).
\end{align}
To derive these equations, we have used the following transformation
\begin{align}
\left(
\begin{array}{c}
x \\
y \\
z
\end{array}
\right)
= \left(
\begin{array}{ccc}
-1 / \sqrt{6} & -1 / \sqrt{6} & 2 / \sqrt{6} \\
1 / \sqrt{2} & - 1 / \sqrt{2} & 0 \\
1 / \sqrt{3} & 1 / \sqrt{3} & 1 / \sqrt{3}
\end{array}
\right)
\left(
\begin{array}{c}
X \\
Y \\
Z
\end{array}
\right),
\end{align}
and the displacement vector $\bu$ follows the same transformation.
To simplify the spin--strain interaction Hamiltonian in Eq.~(\ref{eqn:Hep}), we introduce the
following constraints to the strain tensors in the cubic crystal coordinate:
$\varepsilon_{XX} = \varepsilon_{YY}$ and $\varepsilon_{YZ} = \varepsilon_{ZX}$.
This indicates that the lattice deformations are limited in the plane including both $[001]$ and
$[110]$ axes.
Consequently, the above strain-dependent couplings in Eq.~(\ref{eqn:Ak}) are rewritten as
\begin{align}
& A_{u, \varepsilon} = g_a \varepsilon_1 + g_{\rm B} \varepsilon_{\rm B},~~
A_{v, \varepsilon} = \frac{1}{\sqrt{3}} ( g_b \varepsilon_{U_1} + g_c \varepsilon_{U_2}  ),
\nonumber \\
& A_{zx, \varepsilon} = \frac{1}{\sqrt{6}} ( 2 g_d \varepsilon_{U_1} - g_e \varepsilon_{U_2} ),~~
A_{xy, \varepsilon} = A_{yz, \varepsilon} = 0,
\end{align}
with the bulk strain $\varepsilon_{\rm B}$ and other strain components
\begin{align}
& \varepsilon_1 = \frac{1}{\sqrt{3}} ( \varepsilon_{YZ} + \varepsilon_{ZX} + \varepsilon_{XY} ),
\nonumber \\
& \varepsilon_{U_1} = \frac{1}{\sqrt{3}} (2\varepsilon_{ZZ} - \varepsilon_{XX} - \varepsilon_{YY}),
\nonumber \\
& \varepsilon_{U_2} = \frac{1}{\sqrt{3}} (2\varepsilon_{XY} - \varepsilon_{YZ} - \varepsilon_{ZX}).
\label{eqn:strain}
\end{align}
We disregard $\varepsilon_{\rm B}$ in $A_{u, \varepsilon}$ and consider the
simplified spin--strain interaction Hamiltonian [Eq.~(1) in the main text] with the five independent
coupling constants redefined as
\begin{align}
& g_a = \frac{2}{3} ( - h_{41} + h_{43} ),~~
g_b = \frac{1}{4} ( h_{15} - \sqrt{2} h_{16} ),~~g_c = \frac{1}{4} ( 2 h_{15} + \sqrt{2} h_{16} ),
\nonumber \\
& g_d = - \frac{1}{4\sqrt{2}} ( h_{25} - \sqrt{2} h_{26} ),
~~g_e = \frac{1}{2\sqrt{2}} ( 2 h_{25} + \sqrt{2} h_{26} ).
\label{eqn:ga-e}
\end{align}
\par

\subsection{Evaluation of spin--strain coupling parameters $g_i$ ($i = a,b,c,d,e$)}
The NV center is a good candidate for investigating the spin--strain couplings in Eq.~(\ref{eqn:Ak})
and the coupling parameters in Eq.~(\ref{eqn:ga-e}) for the $S = 1$ states
with $C_{3v}$ symmetry.
The ratio $A_{zx} / A_v$ is related to the spin--strain coupling parameters as
\begin{align}
\frac{ A_{zx} }{ A_v }
= \frac{ 2 g_d a_{U_1} - g_e a_{U_2} }
{ \sqrt{2} ( g_b a_{U_1} + g_c a_{U_2} ) }
= \frac{ - h_{25} ( a_{U_1} + 2 a_{U_2} ) + \sqrt{2} h_{26} ( a_{U_1} - a_{U_2} ) }
{ h_{15} ( a_{U_1} + 2 a_{U_2} ) - \sqrt{2} h_{16} ( a_{U_1} - a_{U_2} ) },
\label{eqn:AzxAv}
\end{align}
where $a_\lambda$ ($\lambda = 1, U_1, U_2$) represents the amplitude of a time-dependent
oscillating strain field $\varepsilon_\lambda$.
Note that $\varepsilon_\lambda$ is defined as Eq.~(\ref{eqn:strain}), which is one of the linear
combinations of the cubic-frame components of strain tensor.
As discussed in the main text, the ratio $A_{zx} / A_v$ is measurable with the two-phonon
transition probability.
When we choose a single strain component, for instance,
$\varepsilon_{U_1}$, the ratio $g_d / g_b$ is given as $g_d / g_b = A_{zx} / (\sqrt{2} A_v)$.
Similarly, $g_e / g_c$ can be evaluated by choosing $\varepsilon_{U_2}$.
In the presence of two strain components with the amplitudes, for instance, $a_{U_1} = a_{U_2}$,
we obtain the information on the coupling parameters in Eq.~(\ref{eqn:Ak}) as
$h_{25} / h_{15} = - A_{zx} / A_v$ [see Eq.~(\ref{eqn:AzxAv})].
\par

In the same manner, the ratio $A_u / A_v$ is given as
\begin{align}
\frac{ A_u }{ A_v }
= \frac{ \sqrt{3} g_a a_1}{ g_b a_{U_1} + g_c a_{U_2} }
= \frac{8}{\sqrt{3}} \frac{ ( - h_{41} + h_{43} ) a_1 }
{ h_{15} ( a_{U_1} + 2 a_{U_2} ) - \sqrt{2} h_{16} ( a_{U_1} - a_{U_2} ) }.
\label{eqn:AuAv}
\end{align}
When we choose a single strain component $\varepsilon_{XY}$ ($a_1 = a_{U_2} / 2$ and
$a_{U_1} = 0$) [see Eq.~(\ref{eqn:strain})], we obtain $g_a / g_c = 2 A_u / ( \sqrt{3} A_v )$.
If the bulk strain $\varepsilon_{\rm B}$ with the amplitude $a_{\rm B}$ is taken into account as well
as the strain $\varepsilon_1$ in Eq.~(\ref{eqn:AuAv}), $g_a a_1$ is just replaced by
$g_a a_1 + g_{\rm B} a_{\rm B}$, where $g_{\rm B} = ( 2 h_{41} + h_{43} ) / (3 \sqrt{3})$.

\section{Time-Averaged Magnetic Moment Coupled to Oscillating Strain Fields}
In the two-level system based on the $S=1$ spin with $C_{3v}$ symmetry, the magnetic
moment is induced by the singlet ground state $| \psi_1 \rangle$ in an applied magnetic field
($\bh = \gamma_{\rm e} \bH$), whereas there is no contribution from the excited state
$| \psi_2 \rangle$.
The wave functions $| \psi_1 \rangle$ and $| \psi_2 \rangle$ are given as Eqs.~(4) and (5) in the
main text, respectively.
Under a magnetic filed $(h_x, h_y, h_z) = (h \cos \phi, h \sin \phi, 0)$ perpendicular to
$z \parallel [111]$, the matrix form of the spin operator parallel to the field is expressed as
\begin{align}
M_\parallel = S_x \cos \phi + S_y \sin \phi =
\left(
\begin{array}{cc}
\sin 2 \chi & 0 \\
0 & 0
\end{array}
\right),
\label{eqn:Mxy}
\end{align}
using the basis of $\{ | \psi_1 \rangle, | \psi_2 \rangle \}$.
When the ground state is chosen as the initial state and the two states are coupled to oscillating
strain fields, the time-averaged magnetic moment is obtained as
$\bar{M}_\parallel = ( 1 - \bar{P}_{\psi_1 \rightarrow \psi_2} ) \sin 2 \chi$, where
$\bar{P}_{\psi_1 \rightarrow \psi_2}$ is the transition probability and
$\sin 2 \chi = 2 \bar{h} / \sqrt{ 9 + 4 \bar{h}^2 }$ is a function of the magnetic field
$\bar{h} = h / D$ normalized by the uniaxial crystal field ($3D$ is the level splitting for $h = 0$).
At the specific field direction $\phi = \pi / 2$ ($[1 \bar{1} 0]$), the transition probability $\bar{P}$
completely vanishes owing to $A_T = 0$ (see the contour maps in Fig.~1 of the main text).
When $\phi$ is tilted slightly from $\pi / 2$, a finite transverse coupling generates an abrupt
increase in $\bar{P}$ at $\varepsilon_0 / \omega = 1, 2, 3, \cdots$, and it causes a sharp
resonance peak.
Consequently, $\bar{M}_\parallel$ shows an abrupt decrease and shrinks by half.
\par

This magnetoacoustic resonance is dominated by the longitudinal coupling with $A_L$.
When the field direction is fixed at $\phi = \pi / 2$, the similar resonance can be realized by
a hybrid measurement using weak microwave, which has been proposed as photon-assisted
magnetoacoustic resonance in our recent study.
\cite{Koga20}
Let us consider the two-level system coupled to a low-frequency photon field oscillating
along the $z$ axis ($[111]$), and put an additional photon-mediated coupling strength $\Delta$
into the original spin--strain interaction Hamiltonian [Eq.~(7) in the main text] as
\begin{align}
H_{\rm eff} (t) = \frac{1}{2}
\left(
\begin{array}{cc}
- \tilde{\varepsilon}_0 - A_L \cos \omega t & \Delta \sin \chi + A_T \cos \omega t \\
\Delta \sin \chi + A_T^* \cos \omega t & \tilde{\varepsilon}_0 + A_L \cos \omega t
\end{array}
\right).
\label{eqn:Ht}
\end{align}
Here, $\tilde{\varepsilon}_0 = \omega_0 - \omega_l$ ($\omega_0 \equiv \varepsilon_0 / \hbar$)
represents the detuning, where $\omega_l$ is the photon frequency.
This is derived under the rotating wave approximation with respect to $\omega_l$, assuming
$\omega_l \ll \omega$.
In addition, $\Delta$ can be considered as a real number for a sufficiently small photon coupling
compared to the quadrupole--strain coupling with $A_L$ or $A_T$.
\cite{Koga20}
Since $\Delta$ in the off-diagonal component in Eq.~(\ref{eqn:Ht}) represents the photonic
transition, the $\Delta$ term causes the transition between the two states even for $A_T = 0$.
\par

\begin{figure}
\begin{center}
\includegraphics[width=7cm,clip]{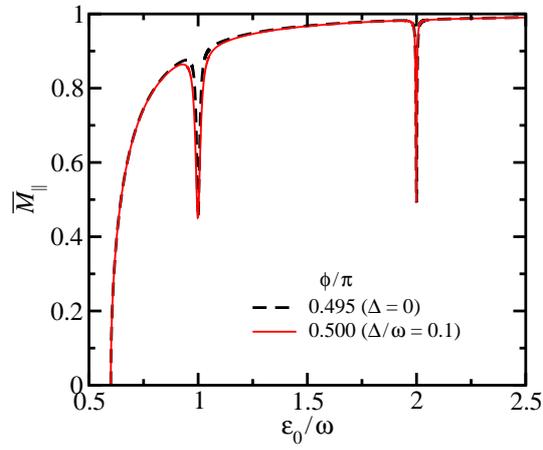}
\end{center}
\caption{
(Color online)
Time-averaged magnetic moment $\bar{M}_\parallel$ induced by a magnetic field in the direction
of $(\cos \phi, \sin \phi, 0)$ (the $z$ axis is chosen as the threefold axis), which is plotted as a
function of $\varepsilon_0 / \omega$ for $D = 0.2$ and
$(A_u, A_v, A_{zx}) = (0, 0.4, 0.4)$ in units of $\omega$.
This is a comparison of the data (black dashed line) for $\phi / \pi = 0.495$ and $\Delta = 0$
(no photon field) with the data (red solid line) for $\phi / \pi = 0.5$ and $\Delta / \omega= 0.1$
(weak photon field).
The energy splitting $\varepsilon_0$ between the two states is related to the magnetic field $h$ as
$\varepsilon_0 / D = [ 3 + \sqrt{ 9 + 4 ( h / D )^2 } ] / 2$.
}
\label{fig:S1}
\end{figure}
In Fig.~S\ref{fig:S1}, we compare the results of $\bar{M}_\parallel$ as a function of
$\varepsilon_0 / \omega$ ($\varepsilon_0 > 3D$):
One is the result for the transverse-phonon coupling effect ($|A_T| \ll |A_L|$, $\Delta = 0$) at
$\phi / \pi = 0.495$ and the other is that for the photon-assisted longitudinal-phonon coupling effect
($|\Delta| \ll |A_L|$, $A_T = 0$) at $\phi / \pi = 0.5$.
It is evident that both results are almost identical, which indicates the equivalency of the phonon-
and photon-mediated transverse couplings in the transition between the two levels.
Note that the former coupling is induced by an oscillating strain field driven by an acoustic wave
or a mechanical oscillator, whereas the latter is due to an ac magnetic field of a microwave with
a background of a high-frequency strain field.
In Fig.~S\ref{fig:S1}, the abrupt shrinking of $\bar{M}_\parallel$ occurs at
$\varepsilon_0 / \omega = 1, 2, 3, \cdots$, and the dips in the $\bar{M}_\parallel$ curve become
narrower as $\varepsilon_0$ increases.
As mentioned above, $\phi = \pi / 2$ is a specific field direction ($[1 \bar{1} 0]$) associated with the
abrupt change in $\bar{M}_\parallel$ at the discrete values of the magnetic field
$h / \omega = \sqrt{ 1 - 3 D / \omega }$, $\sqrt{ 2 ( 2 - 3 D / \omega ) }$, $\cdots$, and these
values correspond to $\varepsilon_0 / \omega = 1$, $2$, $\cdots$, respectively.
This is a very unique point for the present magnetoacoustic resonance, which also holds for the
strong spin--strain coupling.
\par

\section{Derivation of Spin--Strain Interaction Hamiltonian}
In this section, we derive the complete form of the spin--strain interaction Hamiltonian in
Eqs.~(\ref{eqn:Hep}) and (\ref{eqn:Ak}) for $C_{3v}$ symmetry using a group theoretical
analysis.
\cite{Donoho64,Nye85,Nowick95,Powell10,Matsumoto20}
Let us start from the general expression in the following form:
\begin{align}
H_{\varepsilon} = \sum_{ijkl} K_{ijkl} S_i S_j \varepsilon_{kl}~~~~~~(i, j, k, l = x, y, z).
\label{eqn:HepD}
\end{align}
Here, the strain tensor $\varepsilon_{kl}$ is defined by Eq.~(\ref{eqn:epsilon}),
and $S_i$ is the $i$ component of spin operator.
The coefficient $K_{ijkl}$ represents the coupling of quadrupoles (products of spin operators) and
strains.
The strain tensor is symmetric ($\varepsilon_{kl}=\varepsilon_{lk}$) and there are six degrees of
freedom for $kl$ in $K_{ijkl}$.
Since $H_{\varepsilon}$ is invariant under the time-reversal transformation, $K_{ijkl}$ is real.
To satisfy the Hermitian nature of the Hamiltonian, it must be symmetric ($K_{ijkl}=K_{jikl}$).
This leads to $K_{ijkl} ( S_i S_j + S_j S_i ) = K_{ijkl} O_{ij}$ for $i\neq j$, where $O_{ij}$ is given
by Eq.~(\ref{eqn:Ok}).
Since $O_{ij}$ is symmetric as $\varepsilon_{kl}$, there are six degrees of freedom for $ij$ in
$K_{ijkl}$.
Thus, $K_{ijkl}$ is a real symmetric tensor with respect to both $i\leftrightarrow j$ and
$k\leftrightarrow l$, and it can be reduced to a $6\times 6$ matrix.
\par

The spin--strain interaction Hamiltonian is then rewritten as
\begin{align}
H_{\varepsilon} = \sum_{m,n=xx,yy,zz,yz,zx,xy} \tilde{K}_{mn} \tilde{O}_m \tilde{\varepsilon}_n.
\label{eqn:HepD-2}
\end{align}
Here, the vectorial components with tilde are defined by the six tensorial components as
\cite{Nowick95}
\begin{align}
& \tilde{O}_m \longrightarrow
 \left(O_{xx},O_{yy},O_{zz},\sqrt{2}\frac{O_{yz}}{2},\sqrt{2}\frac{O_{zx}}{2},\sqrt{2}\frac{O_{xy}}{2}
 \right),
\nonumber \\
& \tilde{\varepsilon}_n \longrightarrow
 (\varepsilon_{xx},\varepsilon_{yy},\varepsilon_{zz},\sqrt{2}\varepsilon_{yz},\sqrt{2}\varepsilon_{zx},
 \sqrt{2}\varepsilon_{xy}),
\label{eqn:vector}
\end{align}
where $O_{ii} = S_i^2$.
The matrix $\tilde{K}$ in Eq. (\ref{eqn:HepD-2}) is determined so as
to satisfy the invariance of $H_\varepsilon$ under the symmetry transformations.
This type of tensor $\tilde{K}_{mn}$ is known as a fourth-rank matter tensor.
\cite{Nye85,Nowick95,Powell10}
For the $C_{3v}$ point group, it is given by
\cite{Nowick95}
\begin{align}
\tilde{K}_{mn} \longrightarrow
\begin{pmatrix}
K_{11} & K_{12} & K_{13} & \ds{\frac{K_{14}}{\sqrt{2}}} & 0 & 0 \\
\vspace{-0.5cm}~ \\
K_{12} & K_{11} & K_{13} & \ds{-\frac{K_{14}}{\sqrt{2}}} & 0 & 0 \\
K_{31} & K_{31} & K_{33} & 0 & 0 & 0 \\
\ds{\frac{K_{41}}{\sqrt{2}}} & \ds{-\frac{K_{41}}{\sqrt{2}}} & 0 & \ds{\frac{K_{44}}{2}} & 0 & 0 \\
0 & 0 & 0 & 0 & \ds{\frac{K_{44}}{2}} & K_{41} \\
0 & 0 & 0 & 0 & K_{14} & K_{11} - K_{12}
\end{pmatrix},
\label{eqn:K}
\end{align}
using the basis of the vectors in Eq.~(\ref{eqn:vector}).
Substituting Eqs.~(\ref{eqn:vector}) and (\ref{eqn:K}) for Eq.~(\ref{eqn:HepD-2}), we obtain
\begin{align}
& H_\varepsilon = K_{11} ( \varepsilon_{xx} O_{xx} + \varepsilon_{yy} O_{yy} )
 + K_{33} \varepsilon_{zz} O_{zz}
 + K_{12} ( \varepsilon_{yy} O_{xx} + \varepsilon_{xx} O_{yy} )
\nonumber \\
&~~~~~~
 + K_{13} \varepsilon_{zz} ( O_{xx} + O_{yy} ) + K_{31} ( \varepsilon_{xx} + \varepsilon_{yy} ) O_{zz}
 + \frac{ K_{44} }{ 2 } ( \varepsilon_{yz} O_{yz} + \varepsilon_{zx} O_{zx} )
 + ( K_{11} - K_{12} ) \varepsilon_{xy} O_{xy}
\nonumber \\
&~~~~~~
 + K_{14} \varepsilon_{yz} O_v + \frac{ K_{41} }{2} \varepsilon_v O_{yz}
 + K_{14} \varepsilon_{zx} O_{xy} + K_{41} \varepsilon_{xy} O_{zx},
\end{align}
where $\varepsilon_v = \varepsilon_{xx} - \varepsilon_{yy}$.
Using $O_{xx} + O_{yy} + O_{zz} = S(S+1)$ for spin $S$, this is rewritten as
\begin{align}
& H_\varepsilon = \frac{1}{2} ( 2 K_{31} - K_{11} - K_{12} )
 ( \varepsilon_{xx} + \varepsilon_{yy} ) O_{zz} + ( K_{33} - K_{13} ) \varepsilon_{zz} O_{zz}
\nonumber \\
&~~~~~~
 + \frac{1}{2} [ ( K_{11} - K_{12} ) \varepsilon_v + 2 K_{14} \varepsilon_{yz} ] O_v
 + [ ( K_{11} - K_{12} ) \varepsilon_{xy}  + K_{14} \varepsilon_{zx} ] O_{xy}
\nonumber \\
&~~~~~~
 + \frac{1}{2} ( K_{44} \varepsilon_{yz} + K_{41} \varepsilon_v ) O_{yz}
 + \frac{1}{2} ( K_{44} \varepsilon_{zx}  + 2 K_{41} \varepsilon_{xy} ) O_{zx}.
\end{align}
Here, the terms of a common energy shift have been neglected.
Note that the axis vectors of the $xyz$ coordinate are defined as
$\be_x = (1, -1, 0) / \sqrt{2}$, $\be_y = (1, 1, -2) / \sqrt{6}$, and $\be_z = (1, 1, 1) / \sqrt{3}$.
\par

Following Udvarhelyi et al.,
\cite{Udvarhelyi18}
we finally convert the axis vectors as $\be_x \rightarrow \be_y$ and
$\be_y \rightarrow -\be_x$ to compare $H_\varepsilon$ derived here with the Hamiltonian in
Eqs.~(\ref{eqn:Hep}) and (\ref{eqn:Ak}).
The latter was previously derived in Ref.~\ref{ref:Udvarhelyi18}.
Accordingly, the subscripts and signs of the strain tensors and quadrupole operators are replaced
as $yz \rightarrow - zx$, $zx \rightarrow yz$, $xy \rightarrow - xy$, and $v \rightarrow -v$.
For the spin--strain coupling coefficients, $K_{14} \rightarrow K_{15}$ and
$K_{41} \rightarrow K_{51}$.
The obtained Hamiltonian is then rewritten as
\begin{align}
& H_\varepsilon = \frac{1}{2} ( 2 K_{31} - K_{11} - K_{12} )
 ( \varepsilon_{xx} + \varepsilon_{yy} ) O_{zz} + ( K_{33} - K_{13} ) \varepsilon_{zz} O_{zz}
\nonumber \\
&~~~~~~
 + \frac{1}{2} [ ( K_{11} - K_{12} ) \varepsilon_v + 2 K_{15} \varepsilon_{zx} ] O_v
 + [ ( K_{11} - K_{12} ) \varepsilon_{xy}  - K_{15} \varepsilon_{yz} ] O_{xy}
\nonumber \\
&~~~~~~
 + \frac{1}{2} ( K_{44} \varepsilon_{zx} + K_{51} \varepsilon_v ) O_{zx}
 + \frac{1}{2} ( K_{44} \varepsilon_{yz}  - 2 K_{51} \varepsilon_{xy} ) O_{yz}.
\label{eqn:HepK}
\end{align}
As a result, the spin--strain coupling parameters in Eq.~(\ref{eqn:Hep}) are related to those in
Eq.~(\ref{eqn:HepK}) as follows:
\begin{align}
& h_{41} = \frac{1}{2} ( 2 K_{31} - K_{11} - K_{12} ),~~h_{43} = K_{33} - K_{13},
\nonumber \\
& h_{15} = 2 ( K_{11} - K_{12} ),~~h_{16} = - 2 K_{15},
\nonumber \\
& h_{25} = - 2 K_{51},~~h_{26} = K_{44}.
\end{align}
